\documentclass[twocolumn,twocolappendix]{aastex631}
\usepackage{amsmath}
\usepackage{CJKutf8}

\begin{document}

\title{Vortex Avalanches and Collective Motion in Neutron Stars}%\footnote{Released on March, 1st, 2024}}

\author[0000-0002-0799-9027]{I-Kang Liu (\begin{CJK}{UTF8}{bkai}劉翼綱)\end{CJK}}
\affiliation{School of Mathematics, Statistics and Physics, Newcastle University,\\
 Newcastle upon Tyne, NE1 7RU, United Kingdom}

\author[0000-0002-6813-2443]{Andrew W. Baggaley}
\affiliation{School of Mathematics, Statistics and Physics, Newcastle University,\\
 Newcastle upon Tyne, NE1 7RU, United Kingdom}

%\collaboration{20}{(AAS Journals Data Editors)}

\author[0000-0002-4908-7341]{Carlo F. Barenghi}
\affiliation{School of Mathematics, Statistics and Physics, Newcastle University,\\
 Newcastle upon Tyne, NE1 7RU, United Kingdom}

\author[0000-0003-1044-170X]{Toby S. Wood}
\affiliation{School of Mathematics, Statistics and Physics, Newcastle University,\\
 Newcastle upon Tyne, NE1 7RU, United Kingdom}

%% Note that the \and command from previous versions of AASTeX is now
%% depreciated in this version as it is no longer necessary. AASTeX 
%% automatically takes care of all commas and "and"s between authors names.

%% AASTeX 6.31 has the new \collaboration and \nocollaboration commands to
%% provide the collaboration status of a group of authors. These commands 
%% can be used either before or after the list of corresponding authors. The
%% argument for \collaboration is the collaboration identifier. Authors are
%% encouraged to surround collaboration identifiers with ()s. The 
%% \nocollaboration command takes no argument and exists to indicate that
%% the nearby authors are not part of surrounding collaborations.

%% Mark off the abstract in the ``abstract'' environment. 
\begin{abstract}
We simulate the dynamics of about 600 quantum vortices in a spinning-down cylindrical container using a Gross--Pitaevskii model.
For the first time, we find convincing spatial-temporal evidence of avalanching behaviour resulting from vortex depinning and collective motion.
During a typical avalanche, about 10 to 20 vortices exit the container in a short period, producing a glitch in the superfluid angular momentum and a localised void in the vorticity.
After the glitch, vortices continue to depin and circulate around the vorticity void in a similar manner to that seen in previous point-vortex simulations.
We present evidence of collective vortex motion throughout this avalanche process.
We also show that the effective Magnus force can be used to predict when and where avalanches will occur. 
Lastly, we comment on the challenge of extrapolating these results to conditions in real neutron stars,
which contain many orders of magnitude more vortices.
%\footnote{Abstracts for Research Notes of the American Astronomical Society (RNAAS) are limited to 150 words}.  
%If you exceed this length the
%Editorial office will ask you to shorten it.

\end{abstract}

%% Keywords should appear after the \end{abstract} command. 
%% The AAS Journals now uses Unified Astronomy Thesaurus concepts:
%% https://astrothesaurus.org
%% You will be asked to selected these concepts during the submission process
%% but this old "keyword" functionality is maintained in case authors want
%% to include these concepts in their preprints.

\keywords{Neutron Stars (1108) -- Pulsars (1306)  -- Hydrodynamical simulations (767) }%Classical Novae (251) --- Ultraviolet astronomy(1736) --- History of astronomy(1868) --- Interdisciplinary astronomy(804)}

%% From the front matter, we move on to the body of the paper.
%% Sections are demarcated by \section and \subsection, respectively.
%% Observe the use of the LaTeX \label
%% command after the \subsection to give a symbolic KEY to the
%% subsection for cross-referencing in a \ref command.
%% You can use LaTeX's \ref and \label commands to keep track of
%% cross-references to sections, equations, tables, and figures.
%% That way, if you change the order of any elements, LaTeX will
%% automatically renumber them.
%%
%% We recommend that authors also use the natbib \citep
%% and \citet commands to identify citations.  The citations are
%% tied to the reference list via symbolic KEYs. The KEY corresponds
%% to the KEY in the \bibitem in the reference list below. 

\section{Introduction} \label{sec:intro}

Rotational glitches are sudden, spasmodic changes in the rotation rate of a neutron star,
which result in changes to the (otherwise regular) pulses of radiation detected from pulsars.
Glitches are believed to arise from the spontaneous transfer of angular momentum to the star's solid crust from neutron superfluid in its interior,
which generally rotates more rapidly.
The inner part of the crust comprises a lattice of nuclei immersed in a sea of superfluid neutrons and degenerate electrons~\citep{Baym1971}.
The superfluid component cannot rotate in the manner of a classical fluid, and instead contains a multitude (typically $10^{18}$--$10^{20}$) of superfluid vortices,
each carrying a quantum of circulation $2\pi\hbar/m$,
where $\hbar$ is Planck's reduced constant and
$m=2m_n$ is the mass of a neutron Cooper pair.
These vortices pin to the crustal nuclei \citep[e.g.][]{Avogadro2008,Chamel2008}
preventing the superfluid from spinning down at the same rate as the crust,
which is constantly losing angular momentum through electromagnetic braking.
% It is thought that, once the rotational lag between the crust and superfluid \blue{exceeds some threshold}, %becomes too great,
% \blue{many} %a large number of
% vortices spontaneously depin
% and a fraction of the superfluid's angular momentum is suddenly transferred to the crust,
% producing a glitch.
% The observed range of glitch sizes implies that the number of vortices involved ranges from $10^{7}$ to $10^{15}$,
% and that the amount of superfluid involved in the glitch is comparable to that residing in the crust
% \citep[although part of the core may also be involved, e.g.][]{Andersson2012,Gugercinoglu2014,Newton2015,Haskell2018}.
% 
% The leading paradigm for glitch dynamics is the avalanche model~\citep{Melatos2008},
% which is motivated by the close analogy with magnetic flux avalanches that occur in type-II superconductors \citep{Anderson1975,Field1995}.
% In this model,
% \blue{stresses on pinned vortices gradually accumulate until many are close to the threshold for depinning,
% and a single depinning event can trigger a cascade.
% This leads to a state of self-organized criticality~\citep{Bak1987},
% with a power-law distribution of glitch sizes and an exponential distribution of waiting times between glitches, consistent with the majority of pulsar observations \citep{Melatos2008,Fuentes2019}.}
It is thought that, once the rotational lag between the curst and the superfluid exceeds some threshold, many vortices spontaneously depin and a fraction of the superfluid’s angular momentum is suddenly transferred to the crust, producing a glitch. The leading paradigm of this process is the avalanche model~\citep{Melatos2008}, which is motivated by the close analogy with magnetic flux avalanches in Type-II superconductors \citep{Anderson1975,Field1995}. This model predicts self-organised criticality~\citep{Bak1987} with a power-law distribution of glitch sizes and an exponential distribution of waiting times between glitches, consistent with the majority of pulsar observations~\citep{Melatos2008,Fuentes2019}.
The observed range of glitch sizes implies that the number of vortices involved ranges from $10^{7}$ to $10^{15}$,
and that the amount of superfluid involved in the glitch is comparable to that residing in the crust
\citep[although part of the core may also be involved, e.g.][]{Andersson2012,Gugercinoglu2014,Newton2015,Haskell2018}.

The simplest superfluid model that self-consistently describes vortex pinning is the Gross--Pitaevskii (GP) model, and this has previously been used to study rotational glitches in the neutron star crust \citep{Warszawski2011,Warszawski2012,Melatos2015,Lonnborn2019,Verma2022,Poli2023,Shukla2024}.
Given the huge disparity between the length scales of individual vortex cores ($\sim10\,$fm) and the star itself ($\sim10$\,km),
such models can only resolve a much smaller system, and previous studies have been limited to $\lesssim200$ vortices.
Although these models have reproduced some features of pulsar glitches, it is unclear whether the results can be scaled up to the true parameter regime.
Indeed, \citet{Warszawski2011} found that glitch size decreased as they increased the number of vortices, and so their results focussed on simulations with $<100$ vortices.

An alternative approach is the point-vortex (or vortex-filament) model \citep{Howitt2020,Cheunchitra2024},
which tracks only the position of vortices rather than the density and velocity of the superfluid.
This is more computationally efficient and such models have produced glitches in systems of up to 5,000 vortices.
However, the interactions between vortices, and their interactions with pinning sites, cannot be treated fully in such a model, and must instead be parameterised.
Moreover, the dynamics depend qualitatively on the degree of dissipation in the model,
which must be incorporated in an \textit{ad hoc} manner,
and greatly exceeds the dissipation expected in a neutron star.

We study the dynamics of up to $600$ vortices in a two-dimensional GP model,
in the presence of a spinning down crust.
We show that, if the spin down is sufficiently slow, then rotational glitches do occur and are associated with vortex avalanches and collective motions..
We also demonstrate that the results are essentially independent of the degree of dissipation,
provided that this is made sufficiently small.

\section{Gross--Pitaevskii Modeling} \label{sec:model}

\subsection{Numerical Model}

In the GP framework, the superfluid is characterised by a
mean-field wavefunction, $\psi(\mathbf{r},t)$,
which satisfies the (damped) GP equation:
\begin{equation}
    \mathrm{i}\hbar\frac{\partial\psi}{\partial t}=(1- \mathrm{i}\gamma)\left(\hat{H}_\mathrm{GP}-\mu\right)\psi.
\label{eq:dGPE}
\end{equation}
Here $\hat{H}_\mathrm{GP}$ is the Gross--Pitaevskii Hamiltonian
and $\mu$ is the effective chemical potential,
i.e.~the energetic cost of increasing the superfluid density.
We emphasize that $\mu$ should not be equated to the chemical potential of the neutrons themselves;
as we explain below, its physical role is to set the superfluid density,
and hence the neutron coherence length.
In the rotating frame, the Hamiltonian can be expressed as
\begin{equation}
    \hat{H}_\mathrm{GP} = - \frac{\hbar^2}{2m}\nabla^2
      + V(\mathbf{r},t) + g\left|\psi(\mathbf{r},t)\right|^2 - \boldsymbol{\Omega}(t)\cdot\hat{\mathbf{L}},
\end{equation}
where $m=2m_\mathrm{n}$ is the mass of a superconducting particle (i.e.~a neutron Cooper pair),
$V(\mathbf{r},t)$ is an imposed potential,
$g$ measures the self-repulsion of the superfluid,
$\boldsymbol{\Omega}(t)$ is the angular velocity of the reference frame,
and $\hat{\mathbf{L}}$ is the angular momentum operator.
We will adopt Cartesian coordinates $(x,y,z)$ such that $\boldsymbol{\Omega} = \bigl(0,0,\Omega(t)\bigr)$,
and so $\boldsymbol{\Omega}\cdot\hat{\mathbf{L}} = - \mathrm{i}\hbar\,\Omega\,\partial/\partial\phi$,
where $\phi$ is the angular coordinate around the $z$-axis.
We solve Eq.~\eqref{eq:dGPE} numerically in the $xy$-plane,
assuming that $\psi$ has no $z$-dependence;
for details see Sec.~\ref{sec:numerical_setup}. 
This reduction to two dimensions precludes dynamics that is intrinsically three-dimensional, such as vortex reconnections and Kelvin waves \citep[e.g.][]{Barenghi2001,Link2022}. However, the model still self-consistently incorporates pinning and depinning, which are the essential processes in glitches, and allows hundreds of vortices to be simulated in a computationally feasible timescale.
Future studies will be required to quantify the role of three dimensional dynamics in the glitch process.

The number density and superfluid velocity of the superfluid (in the rotating frame) are defined as
\begin{equation}
    n = |\psi|^2 \qquad \mbox{and} \qquad \mathbf{v} = \frac{\hbar}{m}\boldsymbol{\nabla}\text{Im}\{\ln\psi \} - \boldsymbol{\Omega}\times\mathbf{r}.
\end{equation}
In the absence of rotation or any imposed potential, the ground state for this system would have zero velocity and a uniform density $n_b\equiv\mu/g$, thus we take $\mu$ and $n_b$ as characteristic units for energy and density, respectively.
The characteristic length scale is the coherence length, $\xi\equiv\hbar/\sqrt{m\mu}$, which sets the vortex core size,
and the characteristic time scale is $\tau\equiv\hbar/\mu$.

As mentioned earlier, %described later in Sec.~\ref{sec:neutron_parameters},
the radius of a neutron star exceeds $\xi$ by many orders of magnitude,
and so it is impracticable to model the entire crust (and all of its vortices) in a single numerical GP model.
Previous GP models have therefore resorted to modelling, in effect, a tiny neutron star,
containing of order 100 vortices only.
In this work, we take a slightly different approach, and aim to model a small but representative piece of the crust.
In contrast to previous models \citep{Warszawski2011,Melatos2015,Drummond2017,Lonnborn2019}
the dynamics within our numerical domain are therefore assumed to play a negligible role in
the overall balance of angular momentum within the star.
For this reason, we simply impose the rotation rate of the crust, $\Omega(t)$,
without taking account of any transfer of angular momentum between the superfluid in our domain and the crust.
In common with previous models, we assume that the crust is spun down by electromagnetic radiation at a constant rate,
and so
\begin{equation}
    \Omega(t)=\Omega_0-\dot\Omega t
\label{eq:Omega_ramp}
\end{equation}
where the initial value $\Omega_0$ and spin-down rate $\dot{\Omega}$ are (positive) constants.

The imposed potential $V(\mathbf{r},t)$ is a sum of three contributions:
\begin{equation}
    V(\mathbf{r},t) = V_\mathrm{con}(\mathbf{r}) + V_\mathrm{pin}(\mathbf{r}) + V_\mathrm{cen}(\mathbf{r},t)
\end{equation}
where $V_\mathrm{con}$ is a confining potential, $V_\mathrm{pin}$ is a pinning potential,
and $V_\mathrm{cen}$ is a centripetal potential that balances the centrifugal force.
The confining potential is taken to be a hard-wall potential at cylindrical radius $R_\mathrm{con}$,
i.e.~$V_\mathrm{con}(\mathbf{r})=V_{0,\mathrm{con}}\Theta(r-R_\mathrm{con})$,
where $\Theta(x)$ is the Heaviside step function and $r=(x^2+y^2)^{1/2}$.
The pinning potential consists of $N_\mathrm{pin}$ identical circular Gaussian barriers of height $V_0$ and width $w$:
\begin{equation}
    V_\mathrm{pin}(\mathbf{r})=\sum_{j=1}^{N_\mathrm{pin}}V_0 \mathrm{e}^{-[(x-x_j)^2+(y-y_j)^2]/w^2}.
\end{equation}
In all of the results presented later, the pinning site locations, $(x_j,y_j)$,
are arranged in a square lattice with separation $d_p$.
However, taking randomly distributed pinning sites does not qualitatively affect the dynamics.
Finally, the centripetal potential is taken to be $V_\mathrm{cen}(\mathbf{r},t) = \tfrac{1}{2}m\Omega^2(t)r^2$,
so that the superfluid density remains roughly uniform,
with $|\psi|^2 \simeq n_b$, despite the overall rotation.
This approach is consistent with regarding the system as only a small piece of the star's crust,
across which the superfluid density should not vary significantly.
This is in contrast to previous GP models, which have generally used a harmonic confining potential~\citep{Warszawski2011,Warszawski2012,Melatos2015,Lonnborn2019},
resulting in a density that varies significantly across the domain.
Such density variations also affect the vortex core size and dynamical time scales, which we prefer to avoid.

The dimensionless coefficient $\gamma$ in Eq.~(\ref{eq:dGPE}) introduces dissipation into the superfluid dynamics.
It acts to reduce the total free energy, defined as
%It dissipates sound waves and guarantees that the total free energy, defined as
\begin{equation}
  \int\mathrm{d}^2\mathbf{r}\,\psi^\ast\left(\hat{H}_\mathrm{GP} - \mu\right)\psi.
\end{equation}
%decreases monotonically with time.
In the context of ultracold quantum gases, $\gamma$ arises from the interaction between superfluid and normal components at nonzero temperatures,
and typically has a value $\lesssim10^{-3}$ \citep[e.g.][]{Bradley2008, Blakie2008, Rooney2012}.
In the inner crust of a neutron star, the true dissipation mechanisms are more complicated, and $\gamma$ serves only as a crude parameterisation.
Nonetheless, previous studies have typically adopted values $\gamma \geqslant 0.02$ \citep{Warszawski2011,Warszawski2012,Melatos2015,Lonnborn2019}.
We aim to determine how small $\gamma$ must be such that it plays little (if any) role in the qualitative behaviour of the system.

\begin{table*}[th!]
\caption{The dimensionless parameters used in the model and typical orders of magnitude in a neutron star~\citep[e.g.][]{Warszawski2011,Harding2013}.
$R_\mathrm{con}$ is the domain radius;
$w$ and $V_0$ are the width and height of the pinning potential;
$d_p$ is the separation between pinning sites;
$\Omega_0$ is the initial angular velocity;
$\dot{\Omega} = -\mathrm{d}\Omega/\mathrm{d}t$ is the spin-down rate.
}
\centering
\begin{tabular}{l|ccc}
 & Simulation  & Neutron Star & Neutron Star \\
 & (dimensionless) & (dimensional)  & (dimensionless) \\\hline
 $R_\mathrm{con}$ &  230 & $10$\,km  & $10^{18}$ \\
 $w$ & 1 & 10\,fm & $1$ \\
 $V_0$ & 2 & $1$\,MeV & $4.8$ \\
 $d_p$ & 10 & $10$\,fm &  $1$ \\
 $\Omega_0$ & $(2,4)\pi\times10^{-3}$ & $(10^{0},10^3)$\,s$^{-1}$  & $(10^{-21},10^{-18})$ \\
 $\dot{\Omega}$ & $2.5\pi\times(10^{-8},10^{-7})$  &  $(10^{-24},10^{-5})$\,$s^{-2}$  &   $(10^{-65},10^{-46})$ \\
\hline
\end{tabular}

\label{table:parameters}
\end{table*}

\subsection{Numerical Procedure}

\label{sec:numerical_setup}
We numerically solve the two-dimensional damped GP equation in dimensionless form
by scaling the energy, density, length and time by $\mu$, $n_b$, $\xi$ and $\tau$, respectively.
We use 4\textsuperscript{th}-order Runge--Kutta method with a timestep of $10^{-3}\tau$; %$dt=10^{-3}\tau$;
the wavefunction is discretized on a uniform Cartesian grid and spatial derivatives
are evaluated spectrally.
The domain is a square of size $(512\xi)^2$, with $1024^2$ grid points.
The confining potential has height $V_{0,\mathrm{con}}=1000\mu$ and radius $R_\mathrm{con}=230\xi$.
The pinning potential height and width are $V_0=2\mu$ and $w=\xi$,
and the separation between pinning sites is $d_p=10\xi$, creating  $\sim$1,600 pinning sites.

Simulations are prepared by first evolving in imaginary time,
i.e.~by setting $\gamma=0$, $\Omega = \Omega_0$, and replacing $t\rightarrow \mathrm{i}t$ in Eq.~(\ref{eq:dGPE})~\citep[e.g.][]{Modugno2003},
beginning with a random phase at each grid point.
After evolving in imaginary time for a sufficiently long period (typically $>$7,500\,$\tau$),
the wavefunction achieves a quasi-equilibrium in which the superfluid density and angular momentum are essentially steady. This is taken as the initial condition for the damped GP equation.

Each vortex is associated with a quantum of circulation $2\pi\hbar/m$.
In the absence of pinning sites we would therefore expect the initial number of vortices to be roughly
$N_v \simeq \Omega_0 R_\mathrm{con}^2 m/\hbar = (\Omega_0\tau)(R_\mathrm{con}/\xi)^2$,
so that the average rotation rate of the superfluid is roughly $\Omega_0$.
For example, with $R_\mathrm{con} = 230\xi$ and $\Omega_0 = 2\pi\times10^{-3}\tau^{-1}$
we would expect $N_v \simeq 332$.
To study significantly more vortices than previous GP studies we take 
$\Omega_0 = 2\pi\times10^{-3}\tau^{-1}$ or $\Omega_0 = 4\pi\times10^{-3}\tau^{-1}$.
In practice, the presence of pinning sites
means that there are many different quasi-equilibrium states that can be achieved during imaginary time propagation,
and the actual number of vortices can differ from this prediction by up to 20\%.
Even though the number of pinning sites greatly exceeds the number of vortices,
there are usually a small number ($\lesssim 5\%$) of vortices that are unpinned in the initial state
(see the bottom-left inset of Fig.~\ref{fig:1}~(a), for example).
Following \citet{Liu2024}, we define a vortex to be pinned if it is located within $1.25w$ of a pinning site.

\subsection{Numerical vs. Neutron Star Parameters}
\label{sec:NSparameters}
Adopting a characteristic value of $\xi = 10$\,fm for the coherence length in the crust~\citep[e.g.][]{Mendell1991,Seveso2016,Graber2017},
% Mendell 1991, ApJ 380, 515
% Seveso et al. 2016 MNRAS 455, 3952
we find that %$\mu\equiv\hbar^2/m\xi^2\simeq518\,\text{eV}$ and $\tau \equiv \hbar/\mu \simeq 1.3\times10^{-18}$\,s.
\begin{equation}
  \mu \equiv \frac{\hbar^2}{m\xi^2} \simeq 207\,\text{keV}
\end{equation}
and 
\begin{equation}
\tau \equiv \frac{\hbar}{\mu }\simeq 3.2\times10^{-21}\,\textrm{s}.%3.1765e-21
\end{equation}
A  comparison between the parameter values used in our simulations and those estimated for real neutron stars is given in Table ~\ref{table:parameters}.
For computational feasibility, the domain size and angular velocity we use are far from the typical values for a neutron star.
Our objective in the present work is to include as many vortices as feasible to study any resulting collective dynamics, which requires compromise with regard to the other physical parameters.

\section{Spin-down dynamics}
\label{sec:avalanche}

\begin{figure*}[t!]
 \includegraphics[width=1\textwidth]{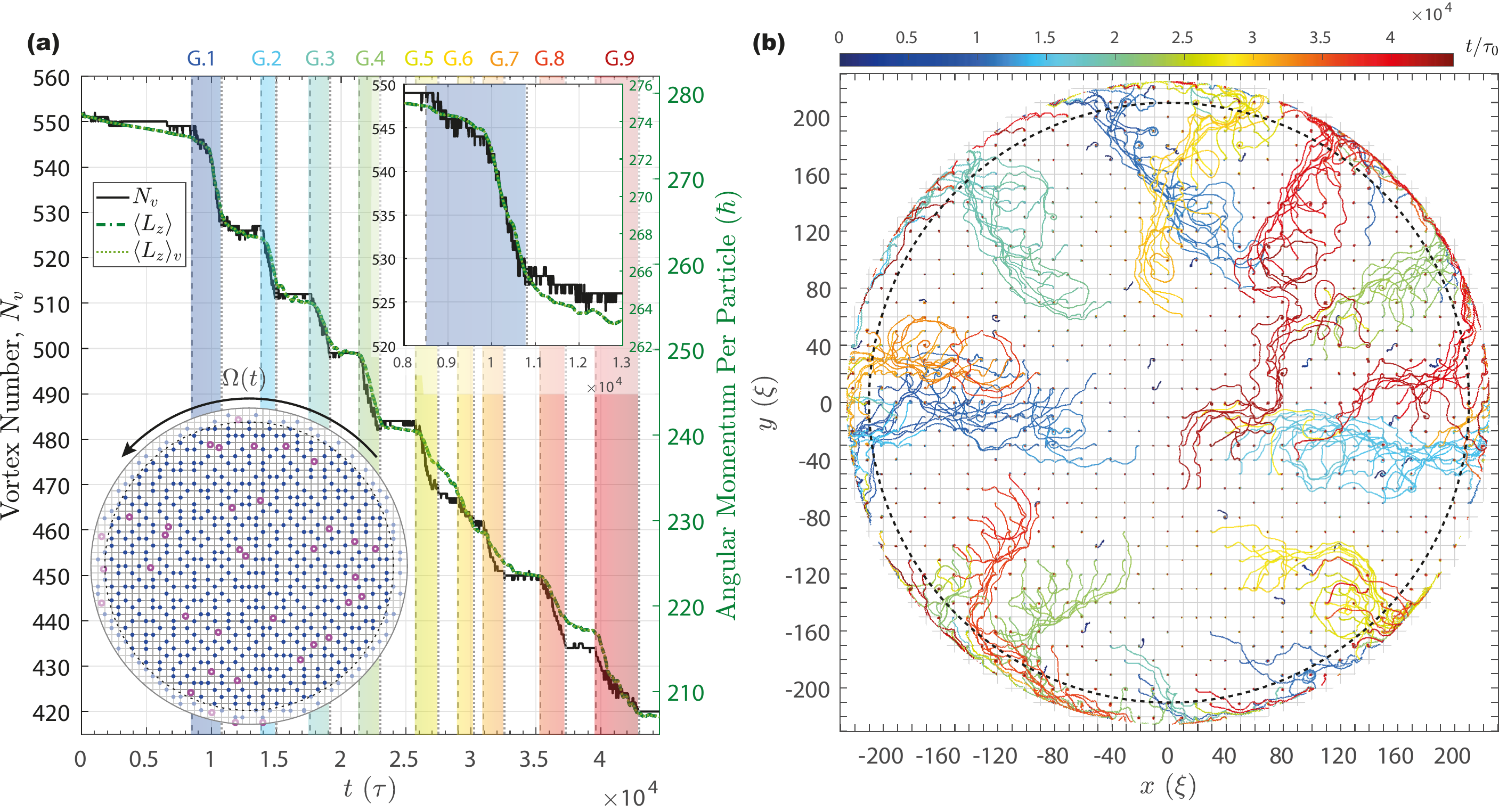}
\caption{
Spin-down dynamics for $\Omega_0=4\pi\times10^{-3}\tau^{-1}$, 
$\dot{\Omega}=2.5\pi\times10^{-8}\tau^{-2}$ and $\gamma=5\times10^{-3}$
over a time span of $4.45\times10^4\tau$.
(a)~Time series of $N_v$ (black solid line), $\langle L_z\rangle$ (green dashed line), and $\langle L_z\rangle_v$ (green dotted line).
Glitches, i.e.~sudden jumps in $\langle L_z\rangle$,
are labelled as G.1 to G.9 and color coded by time.
The bottom-left inset shows the initial vortex locations, whether pinned (blue filled circles) or unpinned (red hollow circles).
The pinning sites are at the vertices of the square grid shown.
The top-right inset shows a close-up of the first glitch~(G.1).
(b)~Vortex trajectories in the rotating frame, colour-coded by time.
The black dashed circle indicates the region $r \leqslant 210\xi$ within which analysis is performed,
i.e.~we take $R = 210\xi$ in Eqs.~\eqref{eq:Lz} and \eqref{eq:Lz_vor}. An animation of the same simulation is included in Appendix~\ref{sec:movie}.
}
\label{fig:1}
\end{figure*}

\subsection{Macroscopic Observables}

We now examine how the initial quasi-equilibrium vortex configuration responds to the linear spin-down imposed by Eq.~\eqref{eq:Omega_ramp}.
As a representative example, we focus on the case with $\gamma=5\times10^{-3}$, $\Omega_0=4\pi\times10^{-3}\tau^{-1}$ and $\dot{\Omega}=2.5\pi\times10^{-8}\tau^{-2}$,
which is illustrated in Fig.~\ref{fig:1}.
In what follows, we will pay particular attention to the number of vortices, $N_v$, and to the mean angular momentum,
\begin{equation}
    \langle L_z \rangle = \left.{\displaystyle\int_{|\mathbf{r}| \leqslant R}\mathrm{d}^2\mathbf{r}\,\psi^\ast\hat{L}_z\psi}\middle/{\displaystyle\int_{|\mathbf{r}| \leqslant R}\mathrm{d}^2\mathbf{r}\,|\psi|^2}\right..
\label{eq:Lz}
\end{equation}
To count the number of vortices we first determine their locations, at a given time,
by interpolating the superfluid velocity to a sub-grid scale and identifying points of singularity.
Identifying vortices is problematic close to the edge of the container, $r=R_\mathrm{con}$, where the density vanishes
and ``ghost vortices'' often arise.
Therefore, our subsequent analysis is performed within the subdomain $|\mathbf{r}|\leqslant R=210\xi\simeq0.91 R_\mathrm{con}$.

Fig.~\ref{fig:1}~(a) presents time-series of $N_v$ and $\langle L_z \rangle$.
The number of vortices exhibits clear step-like drops, each representing a loss of $10$--$25$ vortices from the subdomain within a period of $<3000\tau$.
Each drop is color coded in Fig.~\ref{fig:1}~(a), with dashed and dotted vertical lines indicating its beginning and end time, respectively.
The mean angular momentum exhibits simultaneous step-like behavior, representing a sequence of rotational glitches.

Between glitches, $N_v$ remains essentially constant,
but $\langle L_z\rangle$ decreases smoothly
(at a rate much smaller than $\dot{\Omega}$).
These periods are associated with a spatial redistribution of vortices,
essentially filling in the gaps left by vortices that have left the domain.

The close correlation between $N_v$ and $\langle L_z\rangle$ is not surprising;
 if we neglect variations in the superfluid density, then we can obtain the following estimate for the mean angular momentum within $|\mathbf{r}| \leqslant R$~\citep{Fetter1965}:
\begin{equation}
  \langle L_z \rangle_v = \hbar\sum_{j}\left[1-|\vec{r}_{j}|^2/R^2\right].
\label{eq:Lz_vor}
\end{equation}
Here $\vec{r}_j(t)$ denotes the position of vortex $j$;
for clarity we will use bold typeface for vector fields, and right arrow accents for vector properties of vortices.
As shown in Fig.~\ref{fig:1}~(a),
Eq.~\eqref{eq:Lz_vor} provides an excellent approximation to the true angular momentum, Eq.~\eqref{eq:Lz},
with a relative error of $\simeq 0.5\%$.
From Eq.~\eqref{eq:Lz_vor} we see why the step-like behavior of $N_v$ is also reflected in $\langle L_z\rangle$.
Indeed, as long as the vortices remain roughly uniformly distributed throughout the domain we would expect $\langle L_z\rangle \simeq \tfrac{1}{2}\hbar N_v$.
The slow decreases in $\langle L_z\rangle$ that occurs between glitches must then correspond to a slow outward migration of vortices.

\begin{figure*}[t!]
\begin{center}
\includegraphics[width=0.85\textwidth]{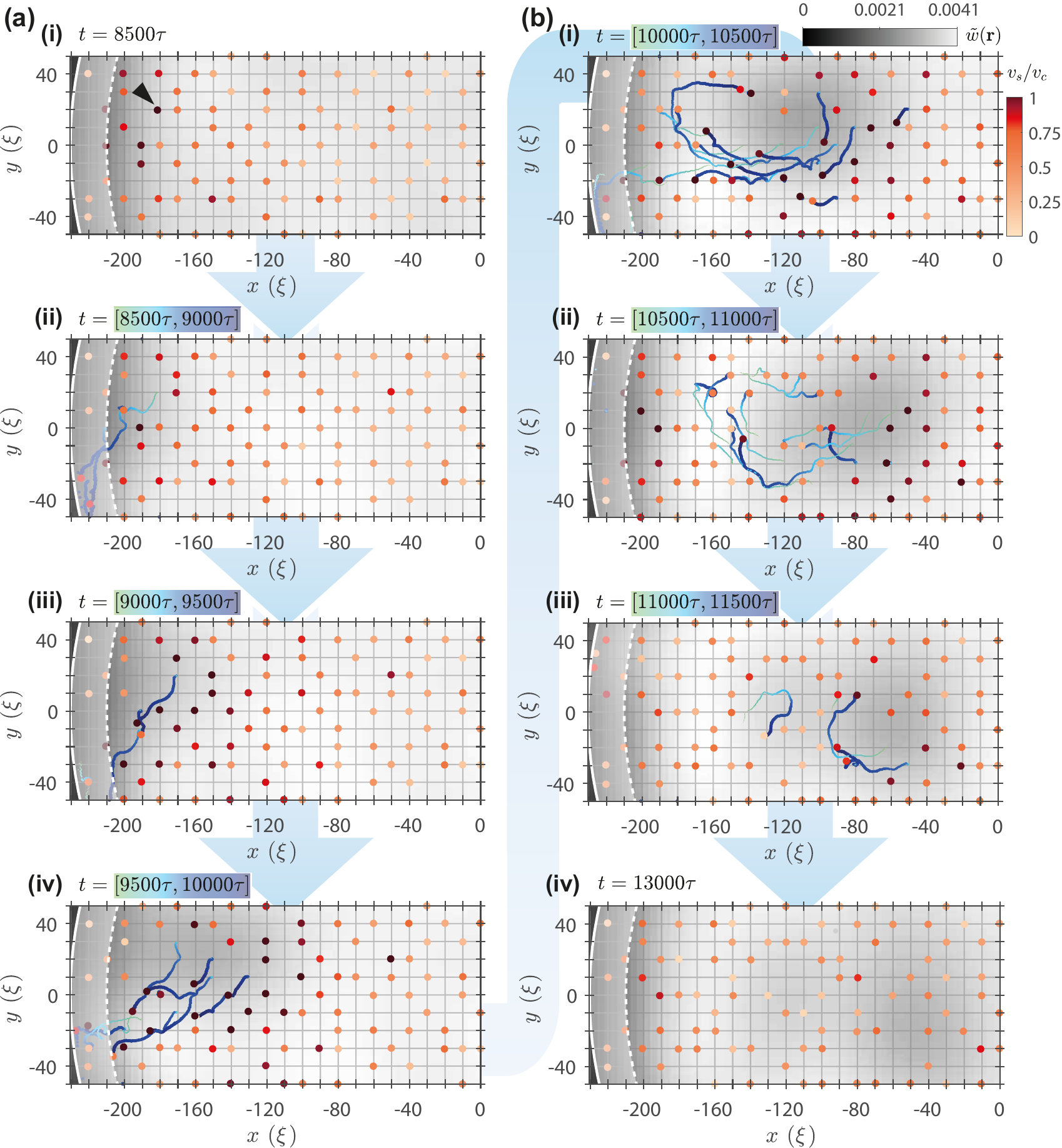}
 \caption{Vortex trajectories for the same simulation shown in Fig.~\ref{fig:1} for four time windows (a) during and (b) after the first glitch~(G.1).
 The locations of vortices at the end of each time window are shown with filled circles, which are color coded to show the magnitude of the Magnus force.
 The trajectories are color coded by time.
 The coarsened vorticity, $\tilde{w}$, is shown in grayscale.
\label{fig:2}}
\end{center}
\end{figure*}

The motion of vortices is illustrated in Fig.~\ref{fig:1}~(b), which plots vortex trajectories
(in the spinning down frame) color coded according to time as in Fig.~\ref{fig:1}~(a).
Each glitch is associated with the unpinning and outward migration of multiple vortices,
localized in both time and space, which we interpret as a \emph{vortex avalanche}.
Each avalanche occurs within a narrow channel that is aligned roughly in the radial direction.
However, individual vortex trajectories are not purely radial, and most follow roughly circular arcs in a clockwise direction, as shown in Fig.~\ref{fig:2}. This glitching behaviour occurs only if the spin-down rate, $\dot{\Omega}$, and dissipation, $\gamma$, are sufficiently small (see Sec.~\ref{sec:ParameterDependence}).

\subsection{Vortex Avalanches}
Several mechanisms can cause a pinned vortex to depin \citep{Warszawski2012},
but the most important for neutron stars is the Magnus force, due to the relative velocity between a pinned vortex and the ambient superfluid flow.
In the case of a point-vortex model, the Magnus force on the $j$-th vortex, with circulation $\vec{\kappa}_j$, is given by
\begin{equation}
  \vec{F}_j = \vec{\kappa}_j\times (\vec{v}_{s,j} - \mathrm{d}\vec{r}_j/\mathrm{d}t),
\end{equation}
where $\vec{v}_{s,j}$ represents the superfluid velocity at the point $\vec{r}_j(t)$.
This superflow is induced by all of the other vortices (plus any image vortices resulting from boundaries):
\begin{equation}
  \displaystyle \vec{v}_{s,j} = \sum_{k\neq j}\frac{\vec{\kappa}_k\times(\vec{r}_j - \vec{r}_k)}{\left|\vec{r}_j - \vec{r}_k\right|^2} - \boldsymbol{\Omega}\times\vec{r}_j.
  \label{eq:vs}
\end{equation}
A vortex is expected to depin when the Magnus force exceeds a critical value.

In practice, the depinning of a vortex in the GP model is more complicated than in this simplified description, and the critical value can vary by about $20\%$ \citep[e.g.][]{Liu2024}.
For the values of $V_0$ and $w$ used here, the critical velocity is $v_c \simeq 0.2\xi/\tau$,
this can help anticipate depinning events.
Fig.~\ref{fig:2} presents snapshots of the vortex locations in the vicinity of the first glitch~(G.1) from Fig.~\ref{fig:1}.
Vortices are coloured according to $|\vec{v}_{s,j}|$ ( Eq.~\eqref{eq:vs}),
where the sum is taken over all vortices in the system.
By $t=\text{8,500}\,\tau$ a few vortices have reached the threshold, $|\vec{v}_{s,j}| > v_c$,
at which depinning is expected;
the first vortex to depin is labelled with a black arrow.
This vortex has a ``knock-on'' collision with another pinned vortex, causing it to depin (see supplementary movie).
As vortices migrate outward, the residual Magnus force on other pinned vortices increases,
as shown in Fig.~\ref{fig:2}(a)(iii), causing further depinning.
The vast majority of depinning events are attributable to the strength of the Magnus force,
with knock-on events being much rarer.

Fig.~\ref{fig:2} also shows the coarsened vorticity of the superfluid, which we define as \citep{Baggaley2012a,Baggaley2012b}
\begin{equation}
  \bar{\omega}(\mathbf{r}) = \sum_{j}^{N_v}\kappa_j W(\left|\vec{r}_{j}-\mathbf{r}\right|,h),
\end{equation}
where $W$ is a smoothing kernel \citep{Monaghan1992} of width $h$.
We choose the value of $h$ based on the average distance between vortices, with
$h(t) = 2.45\sqrt{\pi R_\mathrm{con}^2/N_v(t)}$.
%Fig.~\ref{fig:2} also shows the coarsened vorticity of the superfluid following \citep{Baggaley2012a,Baggaley2012b} using a smoothing kernel with a lengthscale comparable to the separation between vortices.
Figure ~\ref{fig:2}(b) shows the subsequent dynamics, including the post-glitch period.
As a result of vortices exiting the domain, a void appears in the (coarsened) vorticity.
Multiple vortices depin and orbit clockwise around this void, causing the void to propagate inward and gradually disperse.
By $t=\text{13,000}\,\tau$, as shown in Fig.~\ref{fig:2}(b)(iv), all vortices have Magnus forces below the threshold for depinning.
%As stated earlier, this rearrangement of vortices is responsible for the gradual spindown of the superfluid seen between glitch events.
We observe similar dynamics in each of the subsequent glitches, which can be seen in the supplementary movie in Appendix~\ref{sec:movie}.

\section{Vortex Collective Motion} \label{sec:collective}

\begin{figure*}[ht!]
 \includegraphics[width=1\textwidth]{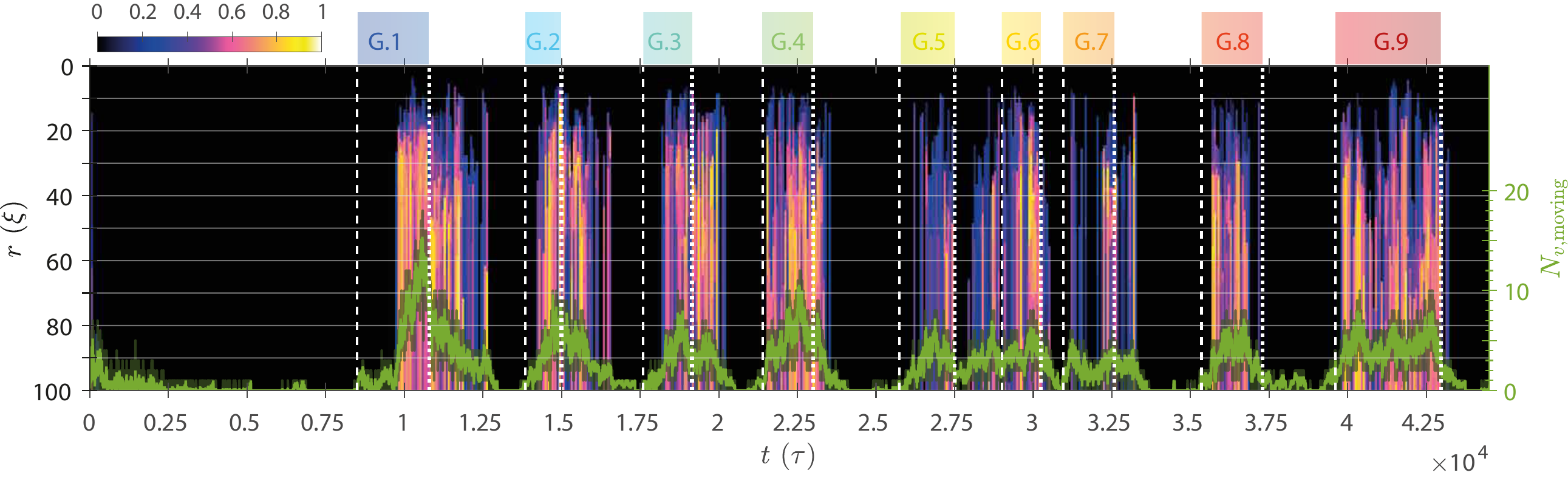}
\caption{The order parameter, $\varphi(r,t)$, (heatmap) averaged over three consecutive measurements,
and the number of moving vortices, $N_{v,\mathrm{moving}}(t)$ (green lines).
The light/dark green lines are smoothed/unsmoothed $N_{v,\mathrm{moving}}$.
The glitches identified in Fig.~\ref{fig:1} are indicated by vertical dashed and dotted lines.
The order parameter vanishes for $r \lesssim 10\xi$, because there are rarely any vortices within this distance,
and saturates for $r \gtrsim 50\xi$, which is the typical size for a cluster of moving vortices.
\label{fig:3}}
\end{figure*}

The vortex trajectories shown in Fig.~\ref{fig:2} suggest that vortices behave collectively both during and after the glitch.  In order to test this, we introduce an order parameter,
\begin{equation}
  \varphi(r,t) = \displaystyle\frac{1}{N_{v,\mathrm{moving}}}\sum_{j=1}^{N_{v,\mathrm{moving}}}\left|\vec{u}_j\right|,
\label{eq:order_parameter}
\end{equation}
where the sum is taken over vortices that are not pinned and are moving,
and where $\vec{u}_j(r,t)$ measures the correlation of all moving vortices within a distance $r$ of vortex $j$, i.e.
\begin{equation}
  \vec{u}_j(r,t) = \frac{1}{\mathcal{N}_j}\sum_{k \in C_j}\frac{\Delta\vec{r}_k/\Delta t}{|\Delta\vec{r}_k/\Delta t|},
\end{equation}
where $C_j = \{k \neq j : |\vec{r}_{j}(t)-\vec{r}_k(t)|\leqslant r\}$, and $\mathcal{N}_j$ is the number of vortices in $C_j$. If $\mathcal{N}_j < 2$ then we omit vortex $j$ from the sum in Eq.~(\ref{eq:order_parameter}). The locations of vortices are tracked every one unit of time, namely, $\Delta t=\tau$, allowing us to identify the vortex motion with sufficiently fine time resolution.

Fig.~\ref{fig:3} presents a time series of $\varphi$ for the same simulation shown in Fig.~\ref{fig:1}.
As in Fig.~\ref{fig:1}, the start and end times of each glitch are indicated by vertical dashed and dotted lines, respectively.
We see that periods of collective motion, indicated by values of $\varphi(r,t) \gtrsim 0.5$ over a range of $r$,
typically occur throughout most of the glitch, and often continue significantly after the glitch.
This is consistent with the dynamics illustrated in Fig.~\ref{fig:2}.
\begin{figure*}[t!]
\begin{center}
%\plotone{fig_5_v5.pdf} 
\includegraphics[width=0.9\textwidth]{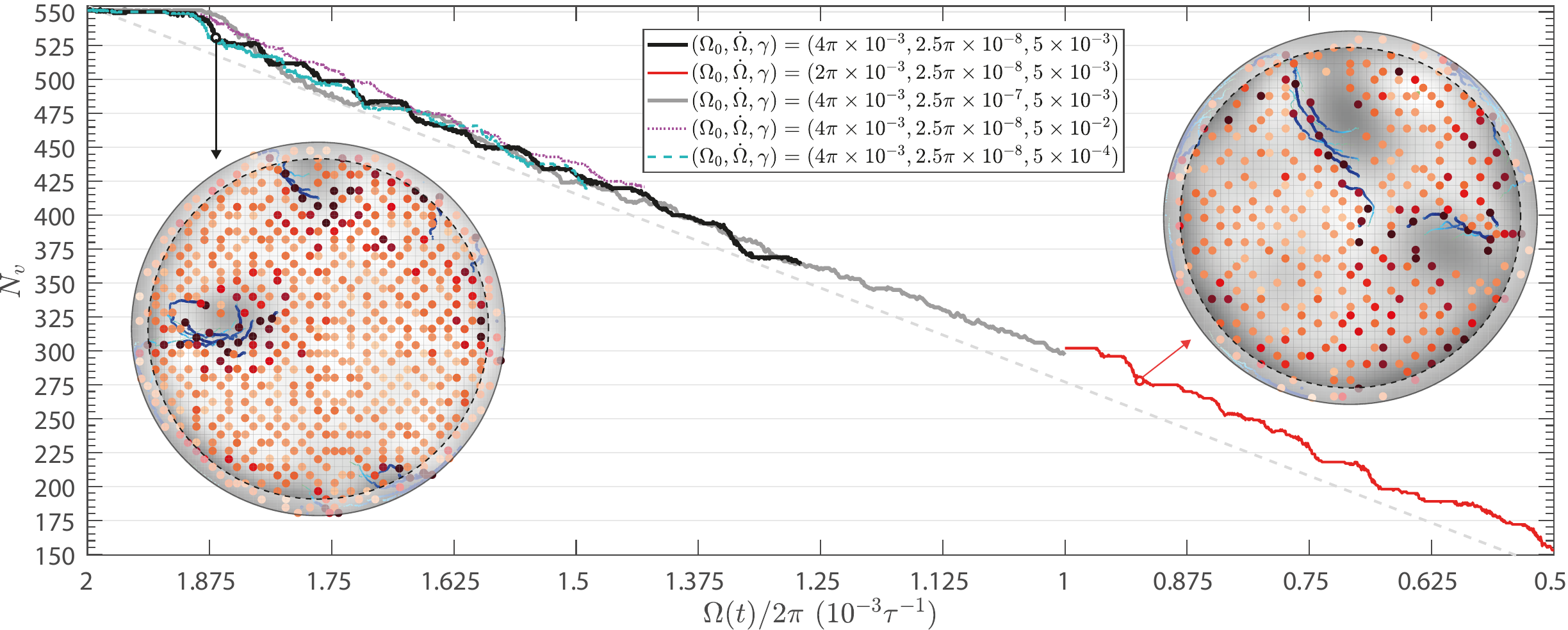}
\caption{Time series of $N_v$ for different simulations.
The solid black line is the representative simulation, as shown in Figs.~\ref{fig:1}--\ref{fig:2}.
The other lines are for faster spin down (solid gray), slower rotation (solid red),
stronger dissipation (dotted purple),
and weaker dissipation (dashed blue). The light grey dashed line indicates the expected value of $N_v$ in the absence of pinning sites.
The insets show the vortex trajectories, Magnus forces and coarsened vorticity immediately after the first glitch in two of the simulations,
using the same color scheme as Fig.~\ref{fig:2}.
\label{fig:4}}
\end{center}
\end{figure*}

\section{ Parameter dependence in GP modeling}
\label{sec:ParameterDependence}
Our results demonstrate the same kind of vortex avalanche behavior believed to occur in neutron stars.
However, given that it is not possible to replicate the true parameter conditions of a neutron star in the computational model,
it is important to determine the extent to which the results depend on the key parameters:
the initial rotation rate, $\Omega_0$, the spin-down rate, $\dot{\Omega}$, and the dissipation, $\gamma$.

\paragraph{Role of initial vortex numbers} 
In the simulation presented in Sec.~\ref{sec:avalanche}, the number of vortices in the region $|\mathbf{r}| \leqslant 210\xi$
decreases from $N_v\simeq550$ to $N_v\simeq420$,
corresponding to only a modest increase in the average distance between vortices,
which remains far smaller than that expected in a neutron star.
To determine whether the same avalanche dynamics persists as the density of vortices decreases,
we halve the initial rotation rate to $\Omega_0 = 2\pi\times10^{-3}\tau^{-1}$.
This simulation initially has around 300 vortices.
As illustrated in Fig.~\ref{fig:4}, we observe similar glitching behavior in both $N_v$ and $\langle L_z\rangle$;
in fact, on average these glitches are larger than those seen in the previous results.
The top-right inset in Fig.~\ref{fig:4} shows the vortex positions at the end of the first glitch in this simulation;
the bottom-left inset presents a similar plot for the previous simulation illustrated in Fig.~\ref{fig:1} and \ref{fig:2}.
We see that, as the average density of vortices decreases,
the vorticity voids produced by each avalanche become larger and more pronounced.

\paragraph{Role of spin-down rate}
Next we investigate the role of spin-down rate by increasing $\dot{\Omega}$ by a factor of 10 to $2.5\pi\times10^{-7}\tau^{-2}$.
As shown in Fig.~\ref{fig:4}, the time series of $N_v$ and $\langle L_z \rangle$ for this simulation
show signs of stochasticity,
but in contrast to the previous results they do not exhibit clear glitching behavior.
Despite this, the vortex trajectories (not shown) display similar patterns to those presented in Figs.~\ref{fig:1} and \ref{fig:2}.
We interpret these results as evidence of multiple vortex avalanches that overlap in time, producing time series that are smoother on average.

\paragraph{Role of dissipation}
Finally, we investigate the role of dissipation by decreasing or increasing $\gamma$ by a factor of 10.
In the simulation with dissipation decreased to $\gamma=5\times10^{-4}$ we observe similar glitching behavior,
and the typical magnitude of the glitches is not significantly affected.
By contrast, in the simulation with dissipation increased to $\gamma=5\times10^{-2}$ we do not observe glitches.
We conclude from this that glitches can only occur when the level of dissipation is sufficiently small,
and that once it is sufficiently small it does not play a significant role in the glitch dynamics.
In our simulations the critical value of $\gamma$ is comparable to the average rotation rate of the superfluid (measured in units of $\tau^{-1}$),
though it likely depends on the rate of spin down, as well as on the choice of pinning potential.

\section{Conclusions} \label{sec:conclusions}
We simulated a rotating superfluid with $\sim 600$ vortices,
coupled to a spinning-down lattice of pinning sites, using a GP model.
For sufficiently slow spin-down, and sufficiently small dissipation,
the vortices undergo avalanches that produce glitches in the superfluid angular momentum.
Each avalanche is triggered when the effective Magnus force on a few neighboring vortices exceeds a critical value, causing depinning.
The movement of these vortices results in stronger Magnus forces on other pinned vortices, producing a cascade of depinning
and creating a localized void in the vorticity.
Depinned vortices circulate anti-cyclonically around this void, which propagates inward and gradually dissipates,
until a new quasi-equilibrium state is achieved.
Throughout this process, the vortex motions are locally correlated, i.e.~they behave collectively.

In a real neutron star the number of vortices, and their mean separation, is many orders of magnitude larger than can be achieved computationally.
However, we have shown that avalanching persists as the mean separation between vortices increases,
provided that the spin-down rate and dissipation are kept sufficiently small.
For more rapid spin-downs, avalanches become so frequent that they overlap in time,
and so the superfluid angular momentum evolves stochastically but without sporadic changes that could be identified as glitches.
A future study may test whether this scenario arises in pulsars, by examining the power series of their spin-down rates.

Our results bear the hallmarks of collective motion, as expected in the standard picture of vortex avalanches and self-organized criticality~\citep{Jensen1998,Melatos2008}.
This collective motion begins during the glitch and often continues into the post-glitch relaxation dynamics.
To determine whether the glitch sizes and waiting times are consistent with the predictions of self-organized criticality theory will require many further simulations,
and will be studied in later work.

Data supporting this publication can be openly accessed under an ‘Open Data Commons Open Database License’ (CC BY 4.0) at {\href{https://doi.org/10.25405/data.ncl.25590531.v1}{https://doi.org/10.25405/data.ncl.25590531.v1}}.
%% IMPORTANT! The old "\acknowledgment" command has be depreciated. It was
%% not robust enough to handle our new dual anonymous review requirements and
%% thus been replaced with the acknowledgment environment. If you try to 
%% compile with \acknowledgment you will get an error print to the screen
%% and in the compiled pdf.
%% 
%% Also note that the akcnowlodgment environment does not support long amounts of text. If you have a lot of people and institutions to acknowledge, do not use this command. Instead, create a new \section{Acknowledgments}.
\begin{acknowledgments}
We thank Brynmor Haskell and Marco Antonelli for fruitful discussions and Vanessa Graber for useful comments.
This work was supported by the Science and Technology Facilities Council grant ST/W001020/1.

\end{acknowledgments}

%\newpage

%\end{document}
%% To help institutions obtain information on the effectiveness of their 
%% telescopes the AAS Journals has created a group of keywords for telescope 
%% facilities.
%
%% Following the acknowledgments section, use the following syntax and the
%% \facility{} or \facilities{} macros to list the keywords of facilities used 
%% in the research for the paper.  Each keyword is check against the master 
%% list during copy editing.  Individual instruments can be provided in 
%% parentheses, after the keyword, but they are not verified.

%\vspace{5mm}
%\facilities{HST(STIS), Swift(XRT and UVOT), AAVSO, CTIO:1.3m,
%CTIO:1.5m,CXO}

%% Similar to \facility{}, there is the optional \software command to allow 
%% authors a place to specify which programs were used during the creation of 
%% the manuscript. Authors should list each code and include either a
%% citation or url to the code inside ()s when available.

%\software{astropy \citep{2013A&A...558A..33A,2018AJ....156..123A},  
%          Cloudy \citep{2013RMxAA..49..137F}, 
%          Source Extractor \citep{1996A&AS..117..393B}
%          }

%% Appendix material should be preceded with a single \appendix command.
%% There should be a \section command for each appendix. Mark appendix
%% subsections with the same markup you use in the main body of the paper.

%% Each Appendix (indicated with \section) will be lettered A, B, C, etc.
%% The equation counter will reset when it encounters the \appendix
%% command and will number appendix equations (A1), (A2), etc. The
%% Figure and Table counter will not reset.

\appendix
\section{Supplementary Movie}\label{sec:movie}
\begin{figure}[h!]
\begin{center}
%\plotone{fig_5_v5.pdf} 
\includegraphics[width=0.475\textwidth]{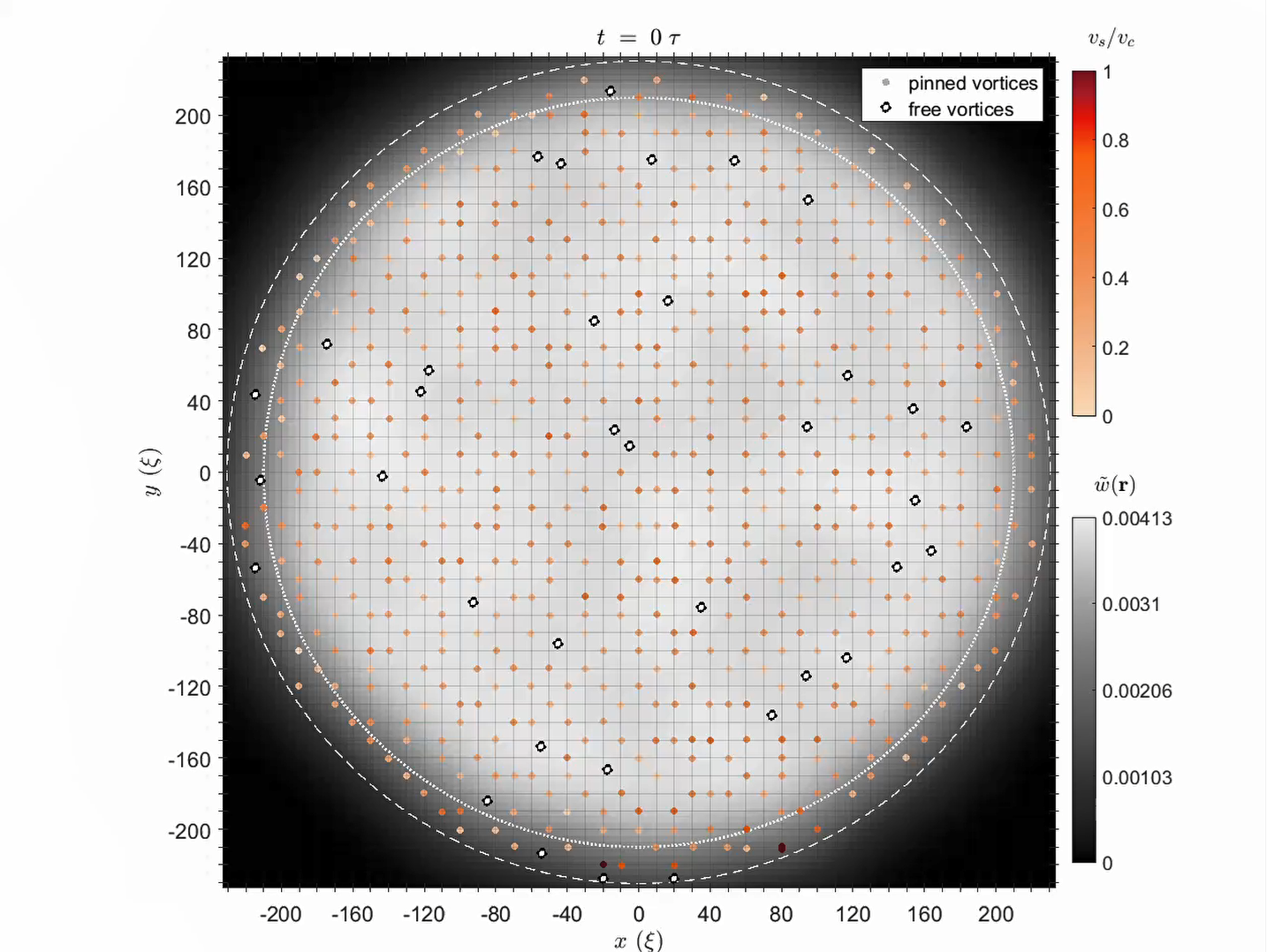}
\caption{Animation of the same simulation shown in Figs.~\ref{fig:1}--\ref{fig:3}.
The color scheme matches that of Fig.~\ref{fig:2}, except that free vortices are marked by hollow black circles.
When the Magnus forces on pinned vortices reaches a threshold, they depin and trigger further depinning, resulting in vortex avalanches. Preview can be seen at \href{https://www.youtube.com/watch?v=dvtMr7haHQo}{https://www.youtube.com/watch?v=dvtMr7haHQo}.
%During and after avalanches vortices move collectively in migrating outward and circulating around voids in the coarsened vorticity, $\tilde{w}$.
\label{fig:5}}
\end{center}
\end{figure}
%\bibliography{ref_ns}{}
\bibliographystyle{aasjournal}

\end{document}